\pdfoutput=1
\documentclass[11pt]{article}

\usepackage[letterpaper,margin=1in]{geometry}

\usepackage[T1]{fontenc}
\usepackage[utf8]{inputenc}
\usepackage{lmodern}
\usepackage{amsmath,amssymb,amsfonts}
\usepackage{graphicx}
\usepackage{booktabs}
\usepackage{array}
\usepackage{tabularx}
\usepackage{enumitem}
\usepackage{authblk}
\usepackage{xcolor}
\usepackage[hidelinks,breaklinks=true]{hyperref}
\usepackage{url}

\title{\bfseries Automated Population-Level Audit Assurance via\\AI-Based Document Intelligence}

\author[1]{Santosh Vasudevan}
\author[2]{Velu Natarajan}
\affil[1]{Caterpillar Inc., Chicago, IL, USA \\ ORCID: 0009-0003-1087-3024}
\affil[2]{GoodRx, Chicago, IL, USA \\ ORCID: 0009-0006-5206-2635}
\date{}

\begin{document}

\maketitle

\begin{center}
\fbox{\parbox{0.92\textwidth}{\centering\small
This paper has been accepted and published at \emph{IEEE SoutheastCon 2026}.
The published version is available via IEEE \emph{Xplore}:\\[0.4em]
IEEE \emph{Xplore}: \href{https://ieeexplore.ieee.org/document/11476237}{https://ieeexplore.ieee.org/document/11476237}\\
DOI: \href{https://doi.org/10.1109/SoutheastCon63549.2026.11476237}{10.1109/SoutheastCon63549.2026.11476237}
}}
\end{center}

\vspace{1em}

\begin{abstract}
\noindent
Audit transaction testing validates accuracy and completeness of customer-facing statements against internal systems of record. Traditional manual, sample-based review of unstructured PDF statements is labor-intensive and does not scale to millions of transactions. This paper presents an automated framework for large-scale audit transaction testing using AI-based document intelligence. The solution leverages Snowflake's Document AI to extract structured data from unstructured PDF statements using a small labeled corpus (approximately 20 documents). Extracted data are reconciled against authoritative source-of-truth datasets to identify discrepancies at scale. Results are surfaced through interactive dashboards and automated reports. The framework enables population-level testing rather than sampling-based approaches, improving audit coverage and supporting continuous assurance objectives. Recent advances in document intelligence have enabled reliable extraction of structured information from complex unstructured documents, while analytics-driven audit automation frameworks have demonstrated potential for scalable and continuous assurance. These technologies support population-level testing that improves audit coverage and enables near real-time risk identification rather than periodic retrospective sampling.

\vspace{0.6em}
\end{abstract}

\section{Introduction}
Audit transaction testing ensures that information recorded in internal systems is accurately reflected in customer-facing communications. In financial services, insurance, and utilities, this includes validating invoices, account summaries, loan disclosures, and confirmations. Regulations including the Sarbanes-Oxley Act (SOX), consumer protection rules from the Consumer Financial Protection Bureau (CFPB), and standards from the Securities and Exchange Commission (SEC) require strong internal controls over financial reporting and customer communications~\cite{sox2002}. Control failures can result in regulatory findings, penalties, and substantial remediation costs~\cite{hammersley2008}.

Traditional transaction testing relies on manual inspection of sampled PDF documents, where auditors compare balances, charges, and dates against internal databases. While effective at small scales, this approach does not scale to millions of statements and introduces inherent sampling risk~\cite{huang2022,chen2022,salijeni2021}. The challenge is amplified by the unstructured nature of customer-facing documents: although source transactions reside in structured systems, final statements are typically generated as unstructured PDFs, creating an automation gap between transaction processing and audit validation workflows~\cite{mahadevkar2024,layoutlmv3,docformer,brownliburd2015}.

Recent advances in document intelligence enable reliable extraction of structured information from complex unstructured documents~\cite{layoutlmv3,docformer}. In parallel, analytics-driven audit frameworks have demonstrated the feasibility of population-level testing and continuous assurance when applied to structured transactional data~\cite{huang2022,chen2022,joshi2020}. However, prior work typically addresses these challenges in isolation---either focusing on structured audit analytics or on document processing without end-to-end integration into the audit transaction testing lifecycle~\cite{mahadevkar2024,salijeni2021}. As a result, fully automated, population-level testing of customer-facing documents remains limited in practice.

This paper addresses these gaps by presenting an AI-driven framework for large-scale audit transaction testing that integrates document intelligence directly with analytic reconciliation and reporting. The proposed solution leverages a cloud-native document intelligence service to extract transactional fields from unstructured customer statements, reconcile extracted values against authoritative source-of-truth tables, and surface discrepancies through interactive dashboards and automated reports~\cite{snowflakedocai}. By shifting from manual, sample-based procedures to scalable automation, the framework enables population-level coverage, improves efficiency, and supports continuous assurance objectives~\cite{huang2022,chen2022,joshi2020}.

The primary contributions of this work are fourfold:
\begin{itemize}[leftmargin=1.4em,itemsep=2pt]
    \item \textbf{End-to-end integration of document intelligence into audit transaction testing} (Section~\ref{sec:arch}): We demonstrate how AI-based document extraction can be operationalized within the audit lifecycle, closing the gap between unstructured customer communications and structured audit analytics.
    \item \textbf{Population-level testing of customer-facing documents} (Sections~\ref{sec:arch}--\ref{sec:experiments}): Unlike prior sampling-based or structured-only approaches, the proposed framework enables automated testing across entire statement populations, eliminating sampling risk.
    \item \textbf{Confidence-aware exception identification} (Section~\ref{sec:results}): By persisting model-generated confidence scores alongside extracted values, the framework supports risk-based prioritization and auditor judgment rather than binary automation.
    \item \textbf{Practitioner-oriented architecture and reporting} (Sections~\ref{sec:arch}--\ref{sec:results}): The solution is designed for deployment within an enterprise data platform, providing transparent, auditable outputs through dashboards and exception reports aligned with audit and compliance workflows.
\end{itemize}

Section~\ref{sec:related} reviews related work. Section~\ref{sec:arch} describes the system architecture and methodology. Section~\ref{sec:experiments} outlines the experimental setup. Section~\ref{sec:results} presents results and testing outcomes. Section~\ref{sec:discussion} discusses implications, and Section~\ref{sec:conclusion} concludes with future directions.

\section{Background and Related Work}
\label{sec:related}

\subsection{Traditional Audit Transaction Testing}
Audit transaction testing has historically relied on manual procedures to validate accuracy and completeness of customer communications. Auditors select samples of statements and compare key fields (balances, amounts, fees, dates) against source-of-truth databases~\cite{huang2022,chen2022,alles2006}. While well-established and aligned with auditing standards, this approach presents limitations:
\begin{itemize}[leftmargin=1.4em,itemsep=2pt]
    \item \textbf{Sampling constraints:} Small subsets of transactions tested, leaving untested populations at residual risk~\cite{huang2022,chen2022}.
    \item \textbf{Manual effort:} Labor-intensive, error-prone extraction of values from unstructured documents~\cite{mahadevkar2024,brownliburd2015,alles2006}.
    \item \textbf{Limited scalability:} Traditional approaches struggle to scale without proportional increases in audit effort~\cite{salijeni2021,alles2006}.
    \item \textbf{Delayed insights:} Testing often occurs after statement issuance, limiting proactive remediation~\cite{salijeni2021,alles2006}.
\end{itemize}

These challenges amplify in environments generating millions of customer documents across billing cycles and regulatory disclosures. Recent work has explored leveraging AI and machine learning to automate audit evidence collection and transaction testing, reducing manual effort and improving coverage~\cite{mahadevkar2024,huang2022,chen2022,salijeni2021}.

\subsection{Regulatory and Compliance Drivers}
Multiple regulatory frameworks emphasize accuracy and integrity of customer communications, increasing the importance of robust testing controls. Key drivers include:
\begin{itemize}[leftmargin=1.4em,itemsep=2pt]
    \item \textbf{Sarbanes-Oxley (SOX):} Requires effective internal controls over financial reporting, ensuring externally communicated information aligns with internal records~\cite{sox2002}.
    \item \textbf{Consumer protection regulations (CFPB):} Mandate accurate, complete, nonmisleading customer statements and disclosures.
    \item \textbf{Industry-specific regulations (banking, insurance, utilities):} Impose requirements for accurate customer communications subject to regulatory review.
\end{itemize}

These regulations increase expectations that organizations demonstrate control effectiveness, including the ability to detect discrepancies between internal systems and customer-facing documents~\cite{hammersley2008}.

\subsection{Automation in Audit and Analytics-Driven Testing}
To address scalability challenges, audit functions have adopted data analytics-driven testing, enabling population-level validation of structured transactional data~\cite{huang2022,chen2022,joshi2020}. These approaches improve coverage and efficiency but typically rely on structured extracts, limiting applicability when audit evidence resides in unstructured PDFs~\cite{mahadevkar2024,salijeni2021}.

Prior work has explored rule-based reconciliation between datasets, statistical anomaly detection, and continuous auditing frameworks~\cite{joshi2020,alles2006}. However, these approaches often exclude unstructured documents or require extensive preprocessing, preventing full automation of end-to-end transaction testing~\cite{mahadevkar2024,brownliburd2015}.

\subsection{Document Intelligence and Unstructured Data Processing}
Recent advances in AI-based document intelligence enable automated extraction of structured information from unstructured documents using machine learning and natural language processing~\cite{mahadevkar2024}. These systems identify fields, interpret layouts, and answer context-aware questions from complex documents including statements, invoices, and forms~\cite{layoutlmv3,docformer,donut}. State-of-the-art models such as DocFormer and OCR-free document understanding transformers have set new benchmarks in document understanding, enabling robust extraction from complex layouts~\cite{docformer,donut}. These advances have been successfully applied in domains such as invoice processing, regulatory compliance, and audit automation~\cite{mahadevkar2024,salijeni2021}. Commercial platforms demonstrate strong performance in invoice processing and form digitization~\cite{snowflakedocai}. Nevertheless, application within audit transaction testing remains limited, particularly for use cases requiring alignment with audit evidence, comparisons against authoritative source systems, and explainable outputs suitable for auditors~\cite{brownliburd2015,alles2006}.

\subsection{Positioning of This Work}
This paper presents a practitioner-oriented framework integrating AI-based document processing directly into the audit transaction testing lifecycle. Unlike traditional sampling-based or structured-only approaches, the proposed solution enables population-level testing of customer-facing documents, supports continuous assurance objectives, and provides transparent outputs through dashboards and reports designed for audit and compliance stakeholders.

\section{System Architecture}
\label{sec:arch}
This section describes the end-to-end architecture of the proposed AI-driven audit transaction testing framework. The workflow consists of three primary stages: document model training, batch inference and data structuring, and analytic validation and reporting. All processing occurs within Snowflake, with user-facing analytics delivered through Streamlit.

\subsection{Platform and Data Environment Setup}
A dedicated Snowflake environment isolates resources for document intelligence analysis. This includes a custom database, schema, and compute warehouse supporting scalable processing of unstructured documents. Raw customer statement PDFs are stored in an encrypted external stage within Snowflake, serving as the centralized repository for documents subject to automated audit testing. Permissions are configured to enable document intelligence capabilities while maintaining governance and security controls consistent with enterprise audit requirements.

\subsection{Document Intelligence Model Training}
A document intelligence model is trained using a small, representative sample of documents (approximately 20 PDFs). For each training document, auditors or subject-matter experts specify the following:
\begin{itemize}[leftmargin=1.4em,itemsep=2pt]
    \item Target fields to extract (e.g., customer identifier, account number, statement balance, minimum payment, due date).
    \item Natural-language prompts describing each field.
    \item Expected ground-truth values for validation.
\end{itemize}

This supervised setup allows the model to learn document layouts, semantic patterns, and contextual cues present in customer statements. Training occurs natively within Snowflake, eliminating need to export sensitive documents to external environments.

\subsection{Scalable Inference and Structured Data Generation}
Once trained, the model is deployed for batch inference across the full population of customer statements in the external stage. During inference, the model processes previously unseen PDFs and predicts values for each target field with associated confidence scores.

Inference is executed at scale by invoking the model's prediction function across the entire document inventory. Secure, pre-signed URLs are generated internally, allowing model access to staged documents without exposing raw files.

The model outputs semi-structured JSON results, subsequently flattened and materialized into a relational table. Each row corresponds to a processed statement and includes:
\begin{itemize}[leftmargin=1.4em,itemsep=2pt]
    \item Extracted field values (customer ID, account number, balance).
    \item Model-generated confidence scores for each field.
    \item Metadata linking results to source documents.
\end{itemize}

This structured representation enables traditional SQL-based validation and anomaly detection workflows familiar to audit practitioners.

\subsection{Transaction Testing and Analytic Reporting}
Structured inference output serves as the foundation for automated transaction testing. Extracted statement values are systematically compared against authoritative internal systems of record including customer master data and transactional databases.

Discrepancies between customer-facing statements and internal records are flagged as potential exceptions. These exceptions, along with confidence scores and metadata, are surfaced to auditors through an interactive Streamlit-based dashboard. The application enables users to:
\begin{itemize}[leftmargin=1.4em,itemsep=2pt]
    \item Filter and prioritize exceptions based on materiality and confidence.
    \item Drill down into individual statements and extracted fields.
    \item Monitor exception trends across customers and reporting periods.
\end{itemize}

This approach shifts transaction testing from manual, sample-based review to continuous, population-level assurance~\cite{huang2022,chen2022,joshi2020,vasarhelyi2004}. Similar architectures integrating document AI with cloud-based analytics platforms have been proposed for large-scale audit and compliance applications~\cite{salijeni2021,alles2006}.

\section{Experimental Setup}
\label{sec:experiments}

\subsection{Dataset and Environment}
The framework was evaluated using customer credit card statements in PDF format. The system processed 500 statements corresponding to 500 unique customers. The trained document intelligence model successfully extracted key financial attributes including minimum payment amount, payment due date, and statement balance from all documents.

This evaluation is a controlled pilot conducted on statements from a single issuer using a largely standardized statement template. As such, the dataset exhibits high consistency in document structure and field semantics, while varying primarily in customer-specific values (e.g., balances, due dates, minimum payments) and minor presentation differences (e.g., spacing and line-wrapping). This scope is representative of many enterprise testing controls deployed per product line or issuer, but it does not fully capture cross-issuer template diversity or highly non-standard document layouts. The statements evaluated in this pilot were primarily digitally generated PDFs; scanned documents were not separately benchmarked.

Implementation was conducted within a Snowflake Standard Edition account on AWS, leveraging native document intelligence capabilities. All data processing, model training, and inference operations were performed using Snowflake SQL and Python UDFs.

\subsection{Model Training and Configuration}
A document intelligence model was trained using 20 representative credit card statement PDFs from historical customer communications. Domain experts defined three primary extraction targets:
\begin{itemize}[leftmargin=1.4em,itemsep=2pt]
    \item \textbf{Minimum Payment Amount:} Lowest payment required to maintain good standing.
    \item \textbf{Payment Due Date:} Deadline for payment receipt.
    \item \textbf{Statement Balance:} Total outstanding account balance.
\end{itemize}

For each field, auditors provided natural-language question prompts (e.g., ``What is the minimum payment due?'') and corresponding ground-truth values.

\subsection{Inference Execution and Data Processing}
Following model training, batch inference was executed across the entire population of customer statements. The inference pipeline included the following steps:
\begin{enumerate}[leftmargin=1.6em,itemsep=2pt]
    \item Enumerated all PDFs in the external stage.
    \item Generated secure, time-limited access URLs for each document.
    \item Invoked the trained model to predict field values and confidence scores.
    \item Persisted results to a Snowflake table in JSON format.
    \item Flattened JSON outputs into relational rows for downstream analysis.
\end{enumerate}

\subsection{Validation Protocol and Metrics}
To improve transparency and reproducibility, extracted fields were evaluated against auditor-confirmed ground truth values for the target attributes (minimum payment amount, payment due date, and statement balance). For each extracted field, we compute precision, recall, and F1-score at the field level. A prediction is counted as a true positive when the extracted value matches the ground truth under a deterministic normalization policy. Normalization includes: (i) currency formatting canonicalization (commas, currency symbols, whitespace), (ii) numeric casting with fixed decimal handling, and (iii) date parsing into a canonical YYYY-MM-DD representation when applicable. Values that are missing, unparsable, or do not match after normalization are counted as errors and routed for exception review.

This protocol reflects how the system would be used in practice: extraction outputs are treated as evidence candidates, reconciled to authoritative records, and any disagreements (including missing values) are surfaced for auditor review rather than silently discarded.

Because the pilot dataset is single-issuer and template-consistent, results should be interpreted as establishing feasibility and operational performance in a controlled setting. In production deployments spanning multiple issuers or evolving layouts, the same framework supports rapid adaptation through (i) small labeled corpus augmentation, (ii) re-training or prompt refinement, and (iii) confidence-aware routing of low-certainty fields to manual review. Broader evaluation across heterogeneous templates, scanned documents, and multilingual statements is an important direction for future work.

\section{Results and Testing Outcomes}
\label{sec:results}

\subsection{Processing Summary}
Across the processed statements, the document intelligence model reported an overall average field-level confidence score of 0.781 at inference time. This aggregate score reflects the model's internal confidence estimates across all extracted fields and is intended to support risk-aware prioritization rather than serve as a direct measure of ground-truth correctness. A breakdown of average confidence scores by extracted attribute reveals variation across field types:
\begin{itemize}[leftmargin=1.4em,itemsep=2pt]
    \item Minimum payment amount: 0.89
    \item Payment due date: 0.675
    \item Statement balance: 0.779
\end{itemize}

These differences are consistent with the relative structural and semantic complexity of each field within unstructured statement layouts. Amount-based fields, such as minimum payment and statement balance, exhibit higher confidence due to consistent formatting, whereas date fields demonstrate lower confidence attributable to layout variability and context-dependent placement.

Subsequent auditor-led manual validation of the extracted fields indicated near-perfect agreement with ground-truth values for the evaluated dataset. Field-level precision and recall for the three target attributes approached unity, with only isolated discrepancies observed. False positive and false negative occurrences were rare and primarily attributable to formatting variability rather than systematic model error. These observations are based on controlled manual validation rather than large-scale benchmark testing and are intended to reflect operational feasibility within a consistent document template.

This assessment included a representative manual validation sample of 20 statements selected across the dataset to verify extraction accuracy against authoritative values.

The observed performance reflects the structural consistency of the single-issuer credit card statement dataset used in this pilot, where document layouts and field semantics are largely standardized across statements. In such settings, document intelligence models are able to leverage stable formatting and contextual cues to achieve high extraction reliability.

This divergence between model-reported confidence scores and observed performance highlights the conservative nature of the Document AI confidence estimates, which are designed to understate certainty and enable practitioners to apply professional judgment when prioritizing review rather than to overstate extraction accuracy.

Table~\ref{tab:baseline} contrasts the traditional manual transaction testing approach with the proposed Document AI-based framework across key audit dimensions.

\begin{table}[ht]
\centering
\caption{Baseline Comparison: Manual Transaction Testing vs.\ Document AI-Based Testing}
\label{tab:baseline}
\renewcommand{\arraystretch}{1.25}
\begin{tabularx}{\textwidth}{l X X}
\toprule
\textbf{Dimension} & \textbf{Manual Approach} & \textbf{Document AI Approach} \\
\midrule
Documents reviewed   & Sample-based (e.g., 0.5\%)   & Full population (100\%) \\
Extraction accuracy  & High (manual)                & 1.00 (validated by audit review) \\
Confidence reporting & Not available                & Model-generated (avg. 0.781) \\
Scalability          & Limited by human effort      & Scales linearly with compute \\
Review latency       & Periodic (batch audits)      & Near real time \\
Sampling risk        & Present                      & Eliminated \\
Audit effort         & High manual effort           & Focused exception review \\
Explainability       & Human judgment               & Field-level values + confidence \\
\bottomrule
\end{tabularx}
\end{table}

\subsection{Exception Identification and Analysis}
Following extraction, structured statement data were programmatically reconciled against internal customer and account master tables. The reconciliation logic evaluated consistency across critical customer-facing fields subject to audit scrutiny:
\begin{itemize}[leftmargin=1.4em,itemsep=2pt]
    \item Minimum payment amount.
    \item Payment due date.
    \item Statement balance.
\end{itemize}

The analysis identified six total potential exceptions across the evaluated population:
\begin{itemize}[leftmargin=1.4em,itemsep=2pt]
    \item Two exceptions related to minimum payment discrepancies.
    \item Two exceptions related to payment due date mismatches.
    \item Two exceptions related to balance inconsistencies.
\end{itemize}

Subsequent auditor review confirmed that all identified potential exceptions represented true discrepancies between customer-facing statements and authoritative internal records. No false positives were observed in the evaluated population, indicating that the automated reconciliation logic effectively isolated materially relevant exceptions without introducing spurious alerts.

These confirmed exceptions represent instances where information communicated to customers did not fully align with internal system records, highlighting potential compliance, operational, or data quality risks. Consistent with prior research, AI-driven document intelligence frameworks have demonstrated strong reliability in extracting audit-relevant information while enabling scalable exception identification and continuous assurance objectives~\cite{mahadevkar2024,huang2022,vasarhelyi2004}.

\subsection{Audit Implications}
Results demonstrate that the proposed framework extends traditional audit transaction testing beyond manual, sample-based review to population-level analysis of unstructured documents. By combining conservative, confidence-scored document intelligence with automated reconciliation and interactive visualization, the approach enables auditors to focus attention on true exceptions rather than routine verification.

The interactive dashboard surfaces extracted values, confidence scores, and identified discrepancies, enabling efficient exception triage, drill-down analysis, and audit evidence generation. This design supports near-real-time identification of customer-facing discrepancies, reduces reliance on manual inspection, and aligns with continuous assurance models adopted in modern audit and compliance functions.

\section{Discussion and Implications}
\label{sec:discussion}

\subsection{Impact on Audit Efficiency and Coverage}
Traditional audit transaction testing relies on manual inspection of limited samples, constrained by time, cost, and human capacity. By automating document ingestion, field extraction, and reconciliation within a cloud-native platform, the proposed workflow enables testing across entire statement populations rather than small samples.

Population-level coverage significantly reduces audit risk by eliminating sampling bias and enabling auditors to focus on true exceptions rather than routine verification. Integration of inference results into interactive dashboards enhances productivity by allowing auditors to quickly identify, filter, and investigate discrepancies~\cite{huang2022,chen2022}.

The shift from manual sampling to automated, population-level testing aligns with industry trends toward continuous auditing and real-time assurance, as demonstrated in recent case studies and frameworks~\cite{joshi2020,salijeni2021,alles2006}.

\subsubsection{Cost-Benefit Analysis and ROI Quantification}
To assess the economic viability of the proposed framework, we conducted a comparative cost analysis between traditional manual audit testing and the automated document intelligence-driven approach. Auditor labor assumptions are based on observed practice, while automation costs reflect conservative estimates derived from publicly available cloud pricing guidance and empirical workload characteristics of the implemented pipeline.

\textbf{Baseline Manual Process Costs.} The baseline audit process consists of three senior auditors compensated at \$85 per hour, reviewing a quarterly sample of 500 statements drawn from a population of approximately 100,000 statements. At an average review time of 15 minutes per statement, the quarterly cost is \$31,875 (500 $\times$ 0.25 hours $\times$ \$85 $\times$ 3 auditors), resulting in an annual expenditure of \$127,500. Under this model, 99.5 percent of the population remains untested, leaving material residual sampling risk.

\textbf{Automated Document Intelligence Costs.} Automation costs are divided into one-time implementation and recurring operational expenses. One-time development costs include document model configuration, dashboard development, and workflow integration, totaling approximately \$9,000. Recurring costs are driven primarily by document inference and lightweight analytical queries. Based on public cloud pricing guidance and conservative assumptions, per-document processing costs are estimated at approximately \$0.03 to \$0.05. At a scale of 100,000 statements per year, recurring annual costs are approximately \$4,000 to \$5,000, inclusive of compute, storage, and dashboard execution.

\begin{table}[ht]
\centering
\caption{Cost Comparison and Savings Over Time}
\label{tab:costs}
\renewcommand{\arraystretch}{1.2}
\begin{tabular}{l r r r r}
\toprule
\textbf{Year} & \textbf{Manual Cost} & \textbf{Automated Cost} & \textbf{Savings} & \textbf{Cumulative} \\
\midrule
Year 0 & \$0       & \$9,000 & $-$\$9,000  & $-$\$9,000  \\
Year 1 & \$127,500 & \$5,000 & \$122,500   & \$113,500   \\
Year 2 & \$127,500 & \$5,000 & \$122,500   & \$236,000   \\
Year 3 & \$127,500 & \$5,000 & \$122,500   & \$358,500   \\
\bottomrule
\end{tabular}
\end{table}

Under these assumptions, the framework achieves a recurring cost reduction exceeding 94 percent relative to manual testing, with payback occurring within the first month of operation. Over a three-year horizon, cumulative savings exceed \$350,000.

Beyond direct financial savings, population-level testing eliminates sampling risk, reduces exception detection latency from quarterly cycles to near-real-time execution, and significantly improves auditor productivity by allowing effort to be redirected from routine verification to higher-value analytical judgment and remediation activities.

\subsection{Reliability and Interpretability of AI-Extracted Data}
A key concern in audit applications of AI is model trustworthiness. The solution addresses this by persisting extracted values alongside associated confidence scores for each field. These measures provide auditors transparency into model reliability and enable risk-based thresholds when identifying exceptions.

The training approach---leveraging a relatively small, curated set of annotated documents---demonstrates that domain-specific document intelligence models generalize effectively to previously unseen documents. This lowers adoption barriers and reduces data-labeling effort typically associated with machine learning initiatives in audit environments~\cite{mahadevkar2024,salijeni2021}.

The importance of model transparency and confidence scoring in audit applications is emphasized in recent literature, which highlights the need for explainable AI in regulatory and assurance contexts~\cite{brownliburd2015,salijeni2021,vasarhelyi2004}.

\subsubsection{Enabling Continuous Assurance}
The proposed architecture naturally supports continuous assurance models. As new customer statements are generated and staged, inference executes automatically, with discrepancies surfaced in near real time. This represents a fundamental shift from periodic, retrospective audits toward ongoing monitoring of customer communications.

Such a model aligns with emerging expectations from regulators and stakeholders for timely detection of errors in customer disclosures. Continuous exception monitoring enables earlier remediation, reducing likelihood of customer harm and regulatory escalation~\cite{joshi2020,alles2006,vasarhelyi2004}.

\subsubsection{Practitioner Adoption Considerations}
From a practitioner perspective, end-to-end implementation within Snowflake, combined with a Streamlit-based user interface, minimizes architectural complexity and supports adoption within existing audit technology stacks. Auditors interact primarily with structured tables, dashboards, and exception reports rather than raw machine learning outputs.

Governance considerations remain critical. Organizations must establish controls around model training, versioning, and access to ensure consistency, auditability, and compliance with internal model risk management standards. Clear documentation of training data, prompts, and evaluation metrics is essential to support reliance on AI-assisted audit procedures.

While the framework demonstrates strong performance within a consistent document template, limitations remain when applied to multilingual statements, non-standard layouts, or heavily scanned documents where OCR artifacts may increase extraction uncertainty. Additionally, model performance may reflect biases present in the training corpus, underscoring the importance of periodic retraining and representative data selection. The system is therefore designed to support, rather than replace, professional auditor judgment through confidence-aware exception routing and human review.

\section{Conclusion and Future Work}
\label{sec:conclusion}
This paper presented an AI-driven framework for automating large-scale audit transaction testing using document intelligence. By leveraging Snowflake's Document AI capabilities to extract structured data from unstructured customer-facing PDF statements and systematically reconciling those outputs against authoritative internal systems of record, the proposed approach addresses a long-standing scalability and efficiency challenge in audit execution. Results demonstrate that transaction testing traditionally performed through manual sampling can be transformed into a repeatable, data-driven process capable of operating at enterprise scale.

The implementation highlights several practical advantages for audit and assurance functions:
\begin{itemize}[leftmargin=1.4em,itemsep=2pt]
    \item \textbf{Platform integration:} Embedding both model training and inference directly within the enterprise data platform reduces architectural complexity and minimizes data movement risks.
    \item \textbf{Confidence-driven triage:} Inclusion of confidence scores alongside extracted fields enables auditors to apply professional judgment by focusing attention on higher-risk exceptions.
    \item \textbf{Interactive reporting:} Integration of dashboards allows audit stakeholders to quickly interpret results, drill into discrepancies, and generate standardized exception reports.
\end{itemize}

From a governance perspective, the framework supports continuous assurance objectives by enabling more frequent and comprehensive testing of customer communications. Rather than relying on periodic, sample-based procedures, audit teams can apply consistent controls across entire document populations, improving coverage while maintaining traceability and auditability of results. This direction aligns with prior work translating continuous auditing theory into operational practice~\cite{alles2008}.

Future work will focus on expanding scope and robustness. Planned extensions include supporting additional document types such as loan statements, insurance policies, and regulatory disclosures; incorporating active learning techniques to continuously improve extraction accuracy; and integrating rule-based and machine-learning-based anomaly scoring to further prioritize exceptions~\cite{joshi2020}. Additional research will explore applicability across multiple regulatory domains and alignment with emerging standards for continuous auditing and assurance.

\section*{Code Availability}
Implementation code is available at: \url{https://github.com/san-to-sh/ieee_doc_ai}.


\begin{thebibliography}{15}

\bibitem{mahadevkar2024}
S.~V. Mahadevkar, S.~Patil, K.~Kotecha, L.~W. Soong, and T.~Choudhury, ``Exploring AI-driven approaches for unstructured document analysis and future horizons,'' \emph{Journal of Big Data}, vol.~11, no.~92, 2024, doi: 10.1186/s40537-024-00948-z.

\bibitem{huang2022}
F.~Huang, W.~G. No, M.~A. Vasarhelyi, and Z.~Yan, ``Audit data analytics, machine learning, and full population testing,'' \emph{Journal of Finance and Data Science}, vol.~8, pp.~138--144, 2022, doi: 10.1016/j.jfds.2022.05.002.

\bibitem{chen2022}
Y.~Chen, Y.~Liu, and J.~Zhang, ``A full population auditing method based on machine learning,'' \emph{Sustainability}, vol.~14, no.~24, Art.\ no.~17008, 2022, doi: 10.3390/su142417008.

\bibitem{joshi2020}
P.~L. Joshi and G.~Marthandan, ``Continuous internal auditing: Can big data analytics help?'' \emph{International Journal of Accounting, Auditing and Performance Evaluation}, vol.~16, no.~1, pp.~25--42, 2020.

\bibitem{layoutlmv3}
Y.~Huang, Y.~Lv, Y.~Cao, L.~Li, and F.~Wei, ``LayoutLMv3: Pre-training for document AI with unified text and image masking,'' in \emph{Proc.\ ACM Int.\ Conf.\ Multimedia (ACM MM)}, 2022.

\bibitem{docformer}
M.~Appalaraju, B.~Jasani, C.~Patel, and B.~Iyer, ``DocFormer: End-to-end transformer for document understanding,'' in \emph{Proc.\ IEEE/CVF Int.\ Conf.\ Computer Vision (ICCV)}, 2021.

\bibitem{sox2002}
\emph{Sarbanes-Oxley Act of 2002}, Pub.\ L.\ No.~107--204, 116 Stat.~745, 2002.

\bibitem{hammersley2008}
J.~S. Hammersley, L.~A. Myers, and C.~Shakespeare, ``Market reactions to the disclosure of internal control weaknesses and the costs of remediation,'' \emph{Journal of Accounting and Economics}, vol.~46, no.~2--3, pp.~291--321, 2008.

\bibitem{brownliburd2015}
H.~Brown-Liburd, H.~Issa, and D.~Lombardi, ``Behavioral implications of big data's impact on audit judgment and decision making,'' \emph{Accounting Horizons}, vol.~29, no.~2, pp.~451--468, 2015.

\bibitem{salijeni2021}
G.~Salijeni, F.~Samsonova-Taddei, and L.~Turley, ``Understanding how big data technologies reconfigure audit practices,'' \emph{Accounting, Auditing \& Accountability Journal}, vol.~34, no.~8, pp.~1751--1778, 2021, doi: 10.1080/09638180.2021.1882320.

\bibitem{snowflakedocai}
Snowflake Inc., ``Snowflake Document AI,'' [Online]. Available: \url{https://docs.snowflake.com/en/user-guide/snowflake-cortex/document-ai/overview}. Accessed: Dec.~06, 2025.

\bibitem{alles2006}
M.~G. Alles, A.~Kogan, and M.~A. Vasarhelyi, ``Continuous monitoring of business process controls: A pilot implementation of a continuous auditing system at Siemens,'' \emph{International Journal of Accounting Information Systems}, vol.~7, no.~2, pp.~137--161, 2006, doi: 10.1016/j.accinf.2005.10.004.

\bibitem{donut}
G.~Kim, T.~Hong, M.~Yim, J.~Nam, J.~Park, J.~Yim, W.~Hwang, S.~Yun, D.~Han, and S.~Park, ``OCR-free document understanding transformer,'' in \emph{Proc.\ Eur.\ Conf.\ Comput.\ Vis.\ (ECCV)}, 2022, pp.~498--517.

\bibitem{vasarhelyi2004}
M.~A. Vasarhelyi, M.~G. Alles, and A.~Kogan, ``Principles of analytic monitoring for continuous assurance,'' \emph{Journal of Emerging Technologies in Accounting}, vol.~1, pp.~1--21, 2004.

\bibitem{alles2008}
M.~G. Alles, A.~Kogan, and M.~A. Vasarhelyi, ``Putting continuous auditing theory into practice: Lessons from two pilot implementations,'' \emph{Journal of Information Systems}, vol.~22, no.~2, pp.~195--214, Fall 2008, doi: 10.2308/jis.2008.22.2.195.

\end{thebibliography}
\end{document}